\documentclass{article}
\usepackage{amsfonts,amssymb, amsmath}
\usepackage[english]{babel}
\usepackage{graphicx}
\usepackage{float}

\def\bq{ \begin{equation}}
\def\eq{ \end{equation}}
\def\ben{ \begin{eqnarray}}
\def\en{ \end{eqnarray}}

\newtheorem{prop}{Proposition}

\begin{document}


\title{Reducible Abelian varieties and Lax matrices for  Euler's problem of two fixed centres}

\author{A.V. Tsiganov \\
\it\small St. Petersburg State University, St. Petersburg, Russia\\
\it\small email: andrey.tsiganov@gmail.com}
\date{}
\maketitle

\begin{abstract}
 Abel's quadratures for integrable Hamiltonian systems are defined up to a group law of the corresponding Abelian variety $A$. If $A$ is isogenous to
 a direct product of Abelian varieties $A\cong A_1\times\cdots\times A_k$, the group law can be used to construct various Lax matrices on the factors $A_1,\ldots,A_k$.
 As an example, we discuss 2-dimensional reducible Abelian variety $A=E_+\times E_-$, which is a product of  1-dimensional varieties $E_\pm$ obtained by Euler in his study of the two fixed centres problem, and the Lax matrices on the factors $E_\pm$.
 \end{abstract}

 \section{Introduction}

 In 1760-1767 Euler considered a point mass moving around the two fixed centers and reduced equations of motion to quadratures, see \cite{eul}.
 Later on, this system attracted the attention of Legendre, Lagrange and Jacobi, who recognized that the solution of the equations of motion can be expressed in terms of elliptic functions and integrals. A long history and a suitable set of references can be found in modern textbooks and papers \cite{bis16,kim18,mar19,mak08}.


 We aim to discuss Abel's approach to Euler's problem. Abel began his Paris memoir \cite{ab} with the observation that “the first idea of [elliptic] functions
was given by the immortal Euler, when he demonstrated that the equation with variables separated
\bq\label{eul-eq1}
\frac{dx}{\sqrt{\alpha+\beta x+\gamma x^2 +\delta x^3 +\epsilon x^4}}+\frac{dy}{\sqrt{\alpha+\beta y+\gamma y^2 +\delta y^3 +\epsilon y^4}}=0
\eq
can be integrated algebraically.”  In \cite{eul} this equation on an elliptic curve was obtained by Euler in his study of the algebraic orbits in
two fixed centers problem as a partial case of equation
\bq\label{eul-eq2}
\frac{dx}{\sqrt{\alpha+\beta x+\gamma x^2 +\delta x^3 +\epsilon x^4}}+\frac{dy}{\sqrt{\alpha'+\beta' y+\gamma' y^2 +\delta' y^3 +\epsilon' y^4}}=0\,,
\eq
on a product of two elliptic curves. When an elliptic curve is realized as a nonsingular cubic curve, its group structure can be described in terms of the sets of three points in which lines intersect the curve, a description that is now well known and widely taught, see Fig.1.
\begin{figure}[H]
\center{\includegraphics[width=0.4\linewidth]{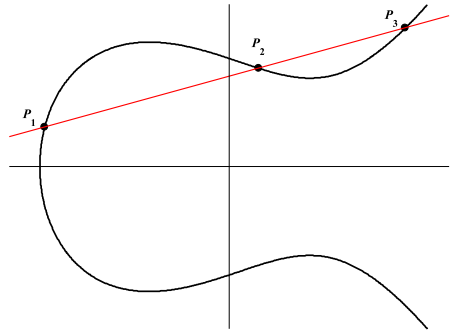} }
\caption{Intersection of an elliptic curve with a line.}
\end{figure}
\par\noindent
There are $N = 3$ intersection points that move with $N-g = 2$ degrees of freedom because only two of the intersection
points can be chosen arbitrarily, where $g=1$ is a genus of the elliptic curve.

A generic approach to the evolution of points was developed by Abel in \cite{ab}, where he sketched a broad generalization of the group
construction. Instead of intersecting a nonsingular cubic curve with an auxiliary line, he intersected an {arbitrary curve} with an {arbitrary family of auxiliary curves}. As the parameters in the defining equation of the auxiliary curve vary, the intersection points vary along the given curve. Abel discovered that, under suitable conditions, $N$ intersection points move in this way with $N- g$ degrees of freedom, where $g$ is the genus of the given curve, see \cite{ed,gg} and references within.

To study equation (\ref{eul-eq2}) we have to consider the intersection of two elliptic curves $E_\pm$ with a line, see Fig.2.
\begin{figure}[!h]
\center{\includegraphics[width=0.4\linewidth]{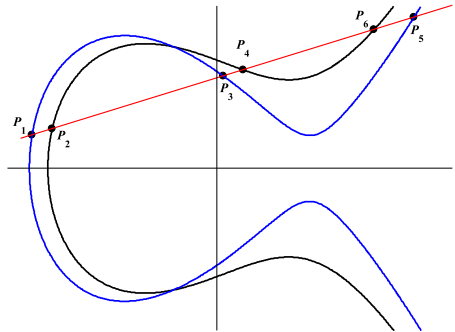} }
\caption{Intersection of two elliptic curves with a line.}
\end{figure}
\par\noindent
As above, there are $N = 6$ intersection points that move with $N-4 = 2$ degrees of freedom because only two of the intersection
points can be chosen arbitrarily.

Thus, we have a 2-dimensional reducible or decomposable Abelian variety  or Abel surface $A$ which is
a direct product  $A= E_+\times E_-$ of one-dimensional Abelian varieties $E_\pm$ (elliptic curves).
Group operations on this reducible Abelian surface $A=E_+\times E_-$ relate all the intersection points, for instance, pairs of the points
\[
(P_i,P_j)\to (P_k,P_m)\,,\qquad i\neq j,\, k\neq m \in \{1,\ldots,6\}\,,
\]
see Fig.2.

In classical mechanics, for integrable by Abel's quadratures dynamical systems, separated variables $x_1,\ldots,x_n$ are affine coordinates of the points $P_1=(x_1,y_1),\ldots, P_n=(x_n,y_n)$ on the Jacobian variety $A=Jac(\Gamma)$ of algebraic curve $\Gamma$, which usually appears as a spectral curve of the Lax matrix $L(x)$. These points $P_k$ are defined up to the group operations on Abelian variety $A$ \cite{fed05,kuz02, ts15a,ts15b,ts15c}. We suppose that this fact is also true for the reducible Jacobian varieties.

In Euler's case, $A=E_+\times E_-$ and we suppose that the following proposition is true.
 \begin{prop}
If affine coordinates of the points $P_i=(x_i,y_i)$ and $P_j=(x_j,y_j)$ in Fig.2. are separated variables for some dynamical system with two degrees of freedom, which satisfy Abel's equations
\[
\frac{dx_i}{\phi_1(x_i)}+\frac{dx_j}{\phi_2(x_j)}=0\,,\qquad
\frac{dx_i}{\phi_3(x_i)}+\frac{dx_j}{\phi_4(x_j)}=dt\,,
\]
 then affine coordinates of another pair of points $P_k=(x_k,y_k)$ and $P_k=(x_k,y_k)$ are also separated variables satisfying Abel's equations
\[
\frac{dx_k}{\varphi_1(x_k)}+\frac{dx_m}{\varphi_2(x_m)}=0\,,\qquad
\frac{dx_k}{\varphi_3(x_k)}+\frac{dx_m}{\varphi_4(x_m)}=dt\,.
\]
Here $\phi_\ell$ and $\varphi_\ell$ are algebraic functions on $x_{i,j}$ and $x_{k,m}$, respectively.
\end{prop}
We have not proofed this proposition in the generic case, see discussion of other partial cases in \cite{bbe86, en96,en18,mag15}.

Our approach is computational. When $A$ is isogenous to a direct product $E_+\times E_-$, group operations on $A$ allow us to construct Mumford's coordinates of the semi-reduced intersection divisors
\[
D_+=P_1+P_3+P_5\qquad\mbox{and}\qquad D_-=P_2+P_4+P_6
\]
on the factors $E_\pm$, which determine $2\times 2$ Lax matrices with an elliptic spectral curves $E_\pm$, see \cite{mum,be90,in07} and references within. Poles of the corresponding Baker-Akhiezer functions are candidates for variables of separations satisfying Abel's equations.

Other Lax matrices appear when we take the hyperelliptic curve of genus two  $\Gamma$ with the Jacobian variety $Jac(\Gamma)$ isogenous to $A= E_+\times E_-$, i.e consider reduction of the hyperelliptic integrals to elliptic ones \cite{hud05,jac32}. With Abel's point of view curve $\Gamma$
\[
\Gamma:\qquad y^2-(a_5x^5+a_4x^4+a_3x^3+a_2x_2+a_1x+a_0)=0\,,
\]
is defined using interpolation by six intersection points $P_1,\ldots,P_n$ up to isogenies. In this case semi-reduced divisor $\widetilde{D}=P_i+P_j$ of degree two generates reduced divisor $\widehat{D}$ in $Jac(\Gamma)$ whose Mumford's coordinates determine $2\times 2$ Lax matrix with the spectral curve $\Gamma$ of genus two \cite{bbe86,bbe94,en96}.

The first example of reducible Abelian variety appears in Euler's two center problem \cite{eul}. Now reducible Abelian varieties became the focus of many mathematicians due to the promising post-quantum cryptography applications, see \cite{bes19,cas96, sas04}. In modern mathematical literature, the key to dealing with algebraic curves is to abandon the notion of points of a curve and to work instead with rational functions on the curve. These rational functions form a field, the algebraic properties of which describe geometric properties of the curve in many cases \cite{ed}. For instance, in Euler's case reducible variety $A= E_+\times E_-$ corresponds to a genus two function field with the elliptic subfields.

The purpose of this note is to come back to Abel's geometric construction, see Fig.2., which allows us to construct a family of $2\times 2$ Lax matrices for Euler's  two center problem. Because of the long history of this problem, it seems unlikely that anything new remains to be discovered. Nonetheless, we have not been able to find similar Lax matrices in the literature.

\section{Reduction of divisors and Lax matrices}
\setcounter{equation}{0}
Let us consider two elliptic curves in the short Weierstrass form
\bq\label{two-ell}
E_{\pm}:\quad y^2-(x^3+a_{\pm}x+b_\pm)=0\,.
\eq
Any pair of points $P_i=(x_i,y_i)$ and $P_j=(x_j,y_j)$ on a direct product of curves $E_+\times E_-$ defines a line
\bq\label{vx}
\Upsilon:\qquad y=V(x)\,,\qquad V(x)=\frac{y_i - y_j}{x_i-x_j}\,x+\frac{x_iy_j -x_jy_i}{x_i-x_j}=v_1x+v_0\,.
\eq
Following to Abel \cite{ab} we substitute $y=V(x)$ into the definitions (\ref{two-ell}) and obtain two Abel's polynomials
\[\Psi_{\pm}(x)=(v_1x+v_0)^2-(x^3+a_{\pm}x+b_{\pm})=0\,.\]
Polynomials $U_{\pm}(x)$
\bq\label{ux}
U_{\pm}(x)=\mbox{MakeMonic}\,\Psi_{\pm}(x)=x^3-v_1^2x^2 +(a_{\pm} - 2v_1v_0)x+b_{\pm}-v_0^2\,,\\
\eq
 are so-called Jacobi polynomials \cite{jac46}, which define abscissas of the intersection points
\[
U_+=(x-x_1)(x-x_2)(x-x_3)\qquad\mbox{and}\qquad U_-=(x-x_2)(x-x_4)(x-x_6)\,.
\]
Two pairs of polynomials $\Bigl(U_+(x),V(x)\Bigr)$ and  $\Bigl(U_-(x),V(x)\Bigr)$ are Mumford's coordinates \cite{mum} of the semi-reduced divisors of degree tree
\[
D_+=P_1+P_3+P_5\qquad\mbox{and}\qquad D_-=P_2+P_4+P_6
\]
on the elliptic curves $E_+$ and $E_-$, see Fig.2.

By definition second Mumford's coordinate $V(x)$ is defined up to the first coordinate $U_\pm(x)$
\[
V(x)\to V_R(x)=V(x)+R_\pm(x)U_\pm(x)
\]
where $R_\pm(x)$ are functions on the phase space, which are arbitrary rational functions on the spectral parameter $x$ without poles in $D_\pm$, respectively.
Thus, we have a family of the Lax matrices depending on $R_\pm(x)$
\bq\label{lax-3}
L_{\pm}(x)=\left(
  \begin{array}{cc}
   V_R(x)& U_{\pm}(x) \\ \\
  W_R&-V_R(x) \\
  \end{array}
 \right)\,,\qquad
   W_R(x)= \dfrac{x^3 + a_{\pm}x + b_{\pm}-V_R^2}{U_{\pm}(x)} \,,
  \eq
with the spectral curves $E_+$ and $E_-$, respectively:
\[
\mbox{det}\Bigl(L_{\pm}(x)-y\Bigr)=y^2-(x^3+a_{\pm}x+b_{\pm})\,.
\]
If $R_\pm(x)$ are polynomials in $x$, then all the entries of $L_\pm(x)$ are polynomials in $x$ including
 $V_R(x)$ and
\[
W_R(x)=1 -U_\pm(x) R_\pm(x)^2 -2 V(x)R_\pm(x)\,.
 \]

 According to Euler \cite{eul} and Lagrange \cite{lag} equations (\ref{eul-eq1},\ref{eul-eq2}) define form of trajectories. In a similar manner Lax matrices $L_{\pm}$ (\ref{lax-3}) allows us to represent the Baker-Akhiezer vector functions $\psi_{\pm}$
 \[
L_\pm(x)\, \psi_\pm(x,y) = y\, \psi_\pm(x,y)
\]
in terms of the Riemann theta function on a nonsingular compactification of the spectral curve.

Time is defined by the additional to (\ref{eul-eq1},\ref{eul-eq2}) equation, for instance
\[
\frac{x^2dx}{\sqrt{\alpha+\beta x+\gamma x^2 +\delta x^3 +\epsilon x^4}}+\frac{y^2dy}{\sqrt{\alpha'+\beta' y+\gamma' y^2 +\delta' y^3 +\epsilon' y^4}}=dt\,.
\]
 Using these equations we can introduce second Lax matrices in the equations
\[
\dfrac{d}{dt} L_\pm=[L_\pm,M_\pm]=L\pm M_\pm-M\pm L_\pm
\]
or in the equations for Baker-Akhiezer vector functions
\[
\frac{d}{dt}\psi_\pm(x,y)=-M_\pm (x)\psi_\pm(x,y)\,.
\]
These matrices are equal to
\bq\label{mlax-3}
M_\pm(x)=\left(
    \begin{array}{cc}
     0 & \dfrac{1}{2V_R}\,\dfrac{d}{dt} U_\pm \\ \\
     \dfrac{1}{U_\pm} \,\dfrac{d}{dt}V_R + \dfrac{W_R}{2V_RU_\pm}\,\dfrac{d}{dt}U_\pm & 0 \\
    \end{array}
   \right)(x)\,,
\eq
see \cite{be90,in07,ts04}.

Poles $P_1,P_2,P_3$ and $P_2,P_4,P_6$ of the corresponding Baker-Akhiezer functions $\psi_{\pm}(x)$
\bq\label{ba-f}
L_{\pm}(x)\, \psi_{\pm}(x,y) = y\, \psi_{\pm}(x,y)
\eq
belong to a line $\Upsilon$ and, therefore, coordinates of the intersection divisors $D_\pm$ satisfy to equations
\bq\label{m-eq}
\frac{dx_1}{y_1}+\frac{dx_3}{y_3}+\frac{dx_5}{y_5}=0\qquad\mbox{and}\qquad
\frac{dx_2}{y_2}+\frac{dx_4}{y_4}+\frac{dx_6}{y_6}=0\,.
\eq
According to Euler \cite{eul}, these equations are reduced to (\ref{eul-eq1}) if one of the points is fixed and $dx_i=0$,
i.e. when equations (\ref{eul-eq1}) have an algebraic Euler's integral. In classical mechanics, this partial case corresponds to the superintegrable systems such that Kepler problem, harmonic oscillator, classical magnets, etc \cite{ts19,ts20r,ts20}.

Let us now suppose that point $P_1=(x_1,y_1)$ belongs to the curve $E_+$ and we know its coordinates. Then polynomials $V(x)=v_1x+v_0$ (\ref{vx}) and
\bq\label{ux-p}
\begin{array}{rcl}
\widetilde{U}_+(x)&=&\dfrac{U_+(x)}{x-x_1}\,\,\mbox{mod}\,\, \{y_1^2= (x_1^3+a_+x_1+b_+)\}\\ \\
&=&x^2+(x_1 -v_1^2)x+a-v_1^2x_1 - 2v_0v_1 + x_1^2\\ \\
&=&(x-x_3)(x-x_5)
\end{array}
\eq
are Mumford's coordinates of the  semi-reduced divisor of degree two on the elliptic curve $E_+$
\[
\widetilde{D}_+=P_3+P_5\,.
\]
Transformation $D_{\pm}\to \widetilde{D}_\pm$ is a standard reduction of divisors on the elliptic curves, which can be performed by using Abel's method or any computer implementation of Cantor's algorithm \cite{bes19}.

Mumford's coordinates of divisor $\widetilde{D}_+$ of degree two determine a family of the $2\times 2$ Lax matrices depending on arbitrary function $R_+(x)$ without poles in $\widetilde{D}$:
\bq\label{lax-2}
\widetilde{L}_+(x)=\left(
  \begin{array}{cc}
   V+R_+\widetilde{U}_+ & \widetilde{U}_+ \\ \\
   \dfrac{x^3 + a_+x + b_+-(V+R_+\widetilde{U}_+)^2}{\widetilde{U}_+} &-V-R_+\widetilde{U}_+ \\
  \end{array}
 \right)\,.
 \eq
The corresponding spectral curve is the elliptic curve $E_+$
\[
\mbox{det}\Bigl(\widetilde{L}_+(x)-y\Bigr)=y^2-(x^3+a_+x+b_+)\,,
\]
and second Lax matrix has the form (\ref{mlax-3}). In similar manner we can get the Lax matrix $\widetilde{L}_-(x)$ with the spectral curve $E_-$.

According to Sklyanin \cite{skl95} Baker-Akhiezer functions $\psi_\pm$ associated with Lax matrices $L_\pm(x)$ and standard normalization
\[ \vec\alpha\cdot \psi_\pm=1\,,\qquad\mbox{where}\qquad \alpha=(0,1)\,,\]
 have three poles on $E_\pm$
 \[P_1,P_3,P_5\qquad\mbox{and}\qquad P_2,P_4,P_6\,. \]
We can remove poles $P_1$ and $P2$ using non-standard "dynamical" normalization
 \[\alpha_+=\left(0,\dfrac{1}{x-x_1}\right)\qquad\mbox{and}\qquad
   \alpha_-=\left(0,\dfrac{1}{x-x_2}\right)\,.
 \]
We can preserve standard normalization $\alpha=(0,1)$ and change Lax matrices
\[
L_\pm(x)\to \widetilde{L}_\pm(x)
\]
using reduction of divisors on hyperelliptic curves and its implementations in the well-studied algorithms and the corresponding software.

\section{Euler's two center problem}
\setcounter{equation}{0}
Let us introduce elliptic coordinates in the orbital plane. If $r_1$ and $r_2$ are distances from a point on the plane to the fixed centers, then elliptic coordinates $u_ {1,2} $ are
\[
r_1+r_2=2u_1\,,\qquad r_1-r_2=2u_2\,.
\]
If the centres are taken to be fixed at $-\kappa$ and $\kappa$ on  $OX$-axis of the Cartesian coordinate system, then we have standard Euler's definition of elliptic coordinates on the plane
\bq
\label{ell-coord}
q_1 =\dfrac{u_1u_2}{\kappa}\qquad \mbox{and}\qquad q_2 = \dfrac{\sqrt{(u_1^2-\kappa^2)(\kappa^2-u_2^2)}}{\kappa}\,.
\eq
Coordinates $u_{1,2}$ are curvilinear orthogonal coordinates, which take values only in the intervals
\[u_2<\kappa<u_1\,,\]
i.e. they are locally defined coordinates. The corresponding momenta are given by
\bq\label{ell-mom}
\begin{array}{rcl}
p_1&=& \dfrac{u_1 u_2 (p_{u_1} u_1-p_{u_2} u_2)-\kappa^2(p_{u_1} u_2-p_{u_2} u_1) }{\kappa(u_1^2- u_2^2)}\,,\\ \\
p_2&=& \dfrac{(p_{u_1} u_1-p_{u_2} u_2)\sqrt{u_1^2-\kappa^2}\sqrt{\kappa^2-u_2^2} }{\kappa(u_1^2-u_2^2)}\,.
\end{array}
\eq
For Euler's  two-centers problem \cite{eul},  in the Cartesian coordinate system Hamiltonian and the first integral are equal to
 \bq\label{int-eul}
 \begin{array}{rcl}
 H_1&=&\dfrac{p_1^2+p_2^2}{2}+\dfrac{\alpha}{2r_1}+\dfrac{\beta}{2r_2} \,,\\
 \\
 H_2&=& (\kappa^2 + q_2^2)p_1^2 - 2q_1q_2p_1p_2 + q_1^2p_2^2
 - \dfrac{\alpha (r_1^2-r_2^2)}{4r_1}-\dfrac{\beta(r_2^2 - r_1^2)}{4r_2}\,.
 \end{array}
 \eq
In elliptic coordinates, these integrals of motion $H_{1,2}$ have the following form
\bq\label{int-eul-e}
 \begin{array}{rcl}
H_1&=& \dfrac{(u_1^2-\kappa^2 )p_{u_1}^2}{2(u_1^2-u_2^2)}
 +\dfrac{(\kappa - u_2^2)p_{u_2}^2}{2(u_1^2-u_2^2)}
 +\dfrac{\alpha}{2(u_1 + u_2)} +\dfrac{\beta}{2(u_1-u_2)}\,,\\
 \\
 H_2&=& \dfrac{u_2^2(u_1^2-\kappa^2)p_{u_1}^2}{u_1^2-u_2^2}
 +\dfrac{u_1^2 (\kappa^2 - u_2^2)p_{u_2}^2}{u_1^2-u_2^2}
 -\dfrac{\alpha u_1u_2}{u_1 + u_2}-\dfrac{\beta u_1 u_2}{u_1-u_2}\,.
 \end{array}
 \eq
Solving the corresponding Hamilton-Jacobi equations $H_{1,2}=h_{1,2}$ with respect to $p_{u_1}$ and $p_{u_2}$ we obtain separated relations
\[
(u_1^2-\kappa^2)^2\, p_{u_1}^2 - (u_1^2-\kappa^2) (2h_1u_1^2 - (\alpha+\beta)u_1-h_2)=0
\]
and
\[
(u_2^2-\kappa^2)^2\,p_{u_2}^2 -(u_2^2-\kappa^2) (2h_1u_2^2 -(\alpha -\beta)u_2 -h_2)=0\,.
\]
Substituting solutions of these equations with respect to $p_{u_1}$ and $p_{u_2}$ into the equations of motion
 \[
 \dfrac{du_1}{dt}=\{u_1,H\}=\dfrac{(u_1^2-\kappa^2)p_{u_1}}{u_1^2-u_2^2}\,,\qquad
 \dfrac{du_2}{dt}=\{u_2,H\}=\dfrac{(u_2^2-\kappa^2)p_{u_2}}{u_2^2-u_1^2}\,,
 \]
 we obtain differential equations of the form
 \[\begin{array}{rcl}
 \dfrac{du_1}{\sqrt{(u_1^2-\kappa^2) (2h_1u_1^2-(\alpha+\beta) u_1-h_2)}}&=&\phantom{-}\dfrac{dt}{u_1^2-u_2^2}\,,\\
 \\
 \dfrac{du_2}{\sqrt{(u_2^2-\kappa^2) (2h_1u_2^2-(\alpha-\beta) u_2-h_2)}}&=&-\dfrac{dt}{u_1^2-u_2^2}\,.
 \end{array}
 \]
The sum of these equations is independent of time and has the form (\ref{eul-eq2})
\bq\label{eq-ab}
 \dfrac{du_1}{\sqrt{(u_1^2-\kappa^2) (2h_1u_1^2-(\alpha+\beta) u_1-h_2)}}+
 \dfrac{du_2}{\sqrt{(u_2^2-\kappa^2) (2h_1u_2^2-(\alpha-\beta) u_2-h_2)}}=0\,.
\eq
According to Euler \cite{eul} and Lagrange \cite{lag} this equation defines a form of trajectories, whereas the second equation
\bq\label{eq-ab-t}
 \dfrac{u_1^2du_1}{\sqrt{(u_1^2-\kappa^2) (2h_1u_1^2-(\alpha+\beta) u_1-h_2)}}+
 \dfrac{u_2^2du_2}{\sqrt{(u_2^2-\kappa^2) (2h_1u_2^2-(\alpha-\beta) u_2-h_2)}}=-dt
\eq
defines time variable. Solutions of these equations are discussed in \cite{bis16}.

In Abel's theory time-independent equation (\ref{eq-ab}) describes the addition of two intersection points on a couple of elliptic curves
\bq\label{epm-4}
E_{\pm}:\quad \nu^2=(\chi^2-\kappa^2) (2h_1\chi^2-(\alpha\pm\beta)\chi-h_2)\,.
\eq
Roughly speaking elliptic coordinates $u_{1,2}$ and the corresponding momenta $p_{u_{1,2}}$
define abscissas and ordinates of the intersection points $P_1$ and $P_2$ on Fig.2:
\bq\label{p12}
P_1=\Bigr(u_1, (u_{1}^2-\kappa^2)p_{u_{1}}\Bigl)\,,\qquad P_2=\Bigr(u_2, (u_{2}^2-\kappa^2)p_{u_{2}}\Bigl)\,.
\eq
Group structure on a real elliptic curve
\[
E:\qquad y^2=a_4x^4+a_3x^3+a_2x^2+a_1x+a_0\,,\qquad a_4\neq 0\,,
\]
 is discussed in \cite{pa99}. Below we prefer to reduce curves (\ref{epm-4})  to the Weierstrass and Legendre forms to use the results of Section 2.

\subsection{Lax matrices with elliptic spectral curves }
In coordinates
\bq\label{trans1}
z=\frac{\chi+\kappa}{\chi-\kappa}\,\Bigl(2h_1\kappa^2 -(\alpha \pm \beta)\kappa - h_2\Bigr)\,,\qquad
y=-\frac{2\kappa \nu}{(\chi-\kappa)^2}(2h_1\kappa^2 -(\alpha\pm\beta)\kappa - h_2)\,.
\eq
equations (\ref{epm-4}) have the form
\bq\label{epm-0}
E_{\pm}:\qquad y^2-z\Bigl(z^2 + 2h_+z +h_-^2-(\alpha\pm\beta)^2\kappa^2\Bigr)=0
\eq
where
\[
h_\pm=2h_1\kappa^2 \pm h_2\,.
\]
After additional transformation
\bq\label{sz-trans}
z= x-\frac{2\,(2h_1\kappa^2 + h_2)}{3}\,,
\eq
equations (\ref{epm-0}) for the curves $E_\pm$ are reduced to the short Weierstrass form (\ref{two-ell})
\bq\label{epm-3}
E_{\pm}:\quad y^2-(x^3+a_{\pm}x+b_{\pm})=0\,.
\eq
where
\bq\label{ab-eul}
\begin{array}{rcl}
a_{\pm}&=&-\kappa^2(\alpha + \beta)^2-\dfrac{(2h_1\kappa^2 - h_2)^2}{3}-\dfrac{32h_1h_2\kappa^2}{3}\,,\\
\\
b_{\pm}&=&\dfrac{2(\kappa^2\Bigl(\alpha \pm\beta)^2 - (4h_1^2\kappa^4 - 68h_1h_2\kappa^2 + h_2^2)/9\Bigr)(2h_1\kappa^2 + h_2)}{3}\,.\\
\end{array}
\eq
Combination of transformations (\ref{trans1}) and (\ref{sz-trans}) has inverse transformation which looks like
\bq\label{sx-trans}
\chi=\frac{ \kappa(3x+2h_1\kappa^2 - 3(\alpha\pm\beta)\kappa - 5h_2 )}{3x-10h_1\kappa^2 + 3(\alpha\pm\beta)\kappa + h_2}\,,\qquad
\nu=\frac{18\kappa(-2h_1\kappa^2 + (\alpha \pm \beta)\kappa + h_2)y }{ (3x-10h_1\kappa^2 + 3(\alpha\pm\beta)\kappa + h_2)^2 }\,.
\eq
Coordinates of the intersection points $P_{1,2}$ (\ref{p12}) on elliptic curves $E_{\pm}$ (\ref{epm-0}) are equal to
\[
z_1=\frac{u_1+\kappa}{u_1-\kappa}\bigl(2h_1\kappa^2 -(\alpha + \beta)\kappa + h_2\bigr)\,,\qquad
z_2=\frac{u_2+\kappa}{u_2-\kappa}\bigl(2h_1\kappa^2 -(\alpha - \beta)\kappa + h_2\bigr)\,.
\]
and
\[
y_1=2\kappa\,p_{u_1}\,z_1\,,\qquad y_2=2\kappa\,p_{u_2}\,z_2\,.
\]
For the elliptic curves $E_{\pm}$ in the Weiershtrass form (\ref{epm-3}), abscissas of the intersection points look like
\bq\label{x12}
x_1=z_1+\frac{2\,(2h_1\kappa^2 + h_2)}{3}\qquad \mbox{and}\qquad
x_2=z_2+\frac{2\,(2h_1\kappa^2 + h_2)}{3}\,.
\eq
Substituting $x_{1,2}$ and $y_{1,2}$ into (\ref{vx}) we obtain Jacobi's polynomials
\[V(x)=\frac{y_i - y_j}{x_i-x_j}\,x+\frac{x_iy_j -x_jy_i}{x_i-x_j}=v_1(x)+v_0\,,
\]
and
\[
U_{\pm}(x)=x^3-v_1^2x^2 +(a_{\pm} - 2v_1v_0)x+b_{\pm}-v_0^2\,,
\]
where $a_\pm$ and $b_\pm$ are given by (\ref{ab-eul}). These Mumford's coordinates of the intersections divisors $D_{\pm}$ define $2\times 2$
Lax matrices $L_\pm(x)$ (\ref{lax-3})
\[
L_{\pm}(x)=\left(
  \begin{array}{cc}
   V_R(x)& U_{\pm}(x) \\ \\
  W_R&-V_R(x) \\
  \end{array}
 \right)\,,\quad V_R(x)=V(x)+R_\pm(x)U_\pm(x)
\]
and Lax matrices $\widetilde{L}_{\pm}(x)$ (\ref{lax-2}).

Explicit expressions of these Lax matrices (\ref{lax-3},\ref{lax-2}) do not interesting to us. We only care about the existence of such matrices and their main property that the genus of the spectral curve det$(L(x)-y)$ is less than the number of degrees of freedom of the corresponding Hamiltonian system.

\subsection{Lax matrices with hyperelliptic spectral curves}
As early as 1832 Legendre had shown that two hyperelliptic integrals are each expressible in terms of two elliptic integrals of the first
kind through a quadratic transformation \cite{leg}. Immediately after, Jacobi \cite{jac32} in a review of Legendre's work pointed out that this property belongs to two linearly independent
integrals of a more general type.

The Jacobi method for the transformation of elliptic integrals can be extended at once to the investigation of the reducibility of hyperelliptic integrals
 to elliptic integrals by a transformation of degree $N$ due to Picard-Weierstrass theorem \cite{hud05,kow84}, see also Konigsberger (1867), Gordan (1869), Pringsheim (1875), Hermite and Caley (1876-77), Picard (1882), Kowalevski (1884), Poincar\'{e} (1884), Bolza (1887), etc. Modern discussion can be found in \cite{cas96, coo84}.

Following \cite{en96}, we consider an inverse problem and introduce Lax matrices with spectral curve $\Gamma$, which is a hyperelliptic curve,
directly starting with Euler's equations (\ref{eq-ab}-\ref{eq-ab-t}). Indeed, let us rewrite equations (\ref{epm-0}) in the standard Legendre form
\bq\label{tori}
E_{\pm}:\qquad\eta^2-\xi(1+\xi)(1+k_{\pm}^2\xi)=0\,,
\eq
with the Jacobi moduli
\[
k_\pm^2=
\frac{h_-^2 -(\alpha \pm \beta)^2\kappa^2}{2h_+^2 - h_-^2 +(\alpha \pm \beta)^2\kappa^2-2h_+\sqrt{h_+^2-h_-^2+(\alpha \pm \beta)^2\kappa^2\,}}\,,
\]
where
\bq\label{trans2}
 \xi=\gamma\, z\,,\qquad\eta=\gamma^{3/2} k_\pm\, y\,,\qquad \gamma=-\frac{(k_\pm^2 + 1)}{2k_\pm^2 h_+}\,.
\eq
Substituting
\[
\xi=\dfrac{\mathcal P(\zeta)}{\mathcal Q(\zeta)}\quad\mbox{and}\quad
\eta=\dfrac{\mathcal P_\pm(\zeta)}{\mathcal Q(\zeta)} \,w
\]
into (\ref{tori}) we obtain definition of the genus $g$ hyperelliptic curve
\[
\Gamma:\qquad w^2=a_{2g+2}\zeta^{2g+2}+a_{2g+1}\zeta^{2g+1}+\cdots+a_0
\]
if polynomials $\mathcal P,\mathcal Q$ and $\mathcal P_\pm$ satisfy the Picard-Weierstrass theorem.

For instance, Jacobi proved that substitution
\bq\label{trans3}
 \xi=\frac{(1-a)(1-b)\zeta}{(\zeta-a)(\zeta-b)}\,,\qquad\eta=-\sqrt{(1-a)(1-b)}\,\frac{\zeta\mp\sqrt{ab}}{(\zeta-a)^2(\zeta-b)^2}\,w\,,
\eq
gives rise to hyperelliptic curve $\Gamma_0$ of genus two
\[
\Gamma_0:\qquad w^2=\zeta(\zeta-1)(\zeta-a)(\zeta-b)(\zeta-ab)\,,
\]
which is a two-sheeted covering of two tori (\ref{tori}) with
\[
k_{\pm}^2=-\frac{(\sqrt{a}\pm\sqrt{b})^2}{(1-a)(1-b)}\,.
\]
Then, applying Rishelot's isogenies $\varrho_k: \Gamma_0\to \Gamma_k$, we can get a tower of hyperelliptic curves
\[
\Gamma_0\to \Gamma_1\to\cdots\to \Gamma_k\to \cdots
\]
 associated with Euler's problem.

 Any hyperelliptic curve $\Gamma_k$ of genus two is a spectral curve of the corresponding $2\times 2$ Lax matrix. Indeed, using a two-sheeted covering of  two tori $\rho_2:\, E_+\times E_-\to \Gamma_0$ and isogeny $\varrho_k $ we can construct a semi-reduced divisor
\[
\widehat{D}_{ij}=(\varrho_k\circ\rho_2)\,(P_i+P_j)\,,
\]
on the genus two hyperelliptic curve $\Gamma_k$. Here $P_i$ and $P_j$ are points on the reducible abelian variety $A=E_+\times E_-$.
Divisor $\widehat{D}_{ij}$ belongs to a class of linearly equivalent divisors which incorporates unique reduced divisor of degree two
\[
\widehat{D}_{ij}\sim \widehat{D}^{r}_{ij}=\widehat{P}_i+\widehat{P}_j\,,
\]
where $\widehat{P}_i$ and $\widehat{P}_j$ are two points on the hyperelliptic curve $\Gamma_k$. This reduced divisor exists according to the Riemann-Roch theorem and its Mumford's coordinates
$\left(\widehat{U}^r,\widehat{V}^r\right)$ can be found by using Cantor's algorithm.

Thus, after transformations (\ref{trans1},\ref{trans2},\ref{trans3}), Richelot's isogeny and reduction of divisors we obtain Mumford's coordinates $\left(\widehat{U}^r,\widehat{V}^r\right)$
of the reduced divisor $\widehat{D}^{r}_{ij}$ on $\Gamma_k$ which is associated with Euler's  equations (\ref{eq-ab}-\ref{eq-ab-t}). The corresponding Lax matrix has the standard form
\[
\widehat{L}(\zeta)=\left(
      \begin{array}{cc}
       \widehat{V}^r & \widehat{U}^r \\
       \widehat{W}^r & -\widehat{V}^r \\
      \end{array}
     \right)(\zeta)\,,\qquad U=(\zeta-\zeta_i)(\zeta-\zeta_j)
\]
 where $\zeta_i$ and $\zeta_j$ are abscissas of two points $\widehat{P}_i$ and $\widehat{P}_j$ on $\Gamma_k$.

The $N$-sheeted covering of two tori $\rho_N:\, E_+\times E_-\to \Gamma$ also generate $2\times 2$ Lax matrices for Euler's problem. When $N>2$ degree of the corresponding reduced divisor
is equal to genus $g$ of the hyperelliptic curve $\Gamma$, which is more than the number degrees of freedom $g>n=2$. It means that the corresponding Baker-Akhieser function has $g>n$ poles and, according to Sklyanin \cite{skl95}, we have a problem with a suitable normalization of the Baker-Akhiezer function.

 As above, we only care about the formal existence of a family Lax matrices for the given Hamiltonian system integrable by Abel's quadratures. Similar $M\times M$ Lax matrices
 with different spectral curves and various numbers of poles of the Baker-Akhiezer function could be obtained by using other methods.

\subsection{Lax matrices in original Cartesian variables}
Let us reconstruct Lax matrices $L_{\pm}$ (\ref{lax-3}) associated with the original Euler's elliptic curves (\ref{epm-4})
\[
E_{\pm}:\quad \nu^2=(\chi^2-\kappa^2) (2h_1\chi^2-(\alpha\pm\beta)\chi-h_2)\,.
\]
According \cite{ab} we consider an intersection of $E_\pm$ with the parabola defining by the equation
\bq\label{par}
\Upsilon:\qquad \nu=V(\chi)\,,\qquad V(\chi)=\sqrt{2h_1}\chi^2+v_1\chi+v_0\,,
\eq
where coefficients $v_1$ and $v_0$ are the following functions on Cartesian variables
\[
v_1=-p_1(\kappa + q_1) - p_2q_2 -\sqrt{2H_1} r_1\,,\qquad v_0=\kappa\left(q_1-\sqrt{2H_1} + p_1r_1\right)\,,
\]
due to the Lagrange interpolation by intersection points
\[
P_1=\Bigl(u_1,p_{u_1}(\kappa^2-u_1^2)\Bigr)\quad\mbox{and}\quad P_2=\Bigl(u_2,p_{u_2}(\kappa^2-u_2^2)\Bigr)\,.
\]
Substituting $\nu=V(\chi)$ into the equations of curves $E_{\pm}$ (\ref{epm-4}) we obtain cubic Abel's polynomials
\bq\label{ux-car}
\begin{array}{rcl}
\Psi_\pm&=&\Psi_\pm^{(3)}\chi^3+\Psi_\pm^{(2)}\chi^2+\Psi_\pm^{1}\chi+\Psi_\pm^{(0)}=
(2v_1\sqrt{2H_1} + \alpha\pm \beta) \chi^3 \\
\\ &+& (2H_1\kappa^2 + 2v_0\sqrt{2H_1} + v_1^2 + H_2)\chi^2 + \bigr(2v_1v_0-\kappa^2(\alpha\pm\beta)\bigl)\chi - \kappa^2 H_1 + v_0^2\,.
\end{array}
\eq
For brevity, we explicitly present only one set of coefficients
\[\begin{array}{rcl}
\Psi_+^{(3)}&=&-\alpha-\frac{ (2r_1-r_2 )\beta}{r_2} - 2(\kappa p_1 + p_1q_1 +p_2q_2)\sqrt{2H_1} -2(p_1^2 + p_2^2)r_1\,,\\
\\
\Psi_+^{(2)}&=&\frac{(4\kappa^2 + 8\kappa q_1 + 3r_1^2 + r_2^2)\alpha}{4r_1}+ \frac{(4\kappa^2 + 8\kappa q_1 + 5r_1^2 - r_2^2)\beta}{4r_2} + 2r_1(2\kappa p_1 + p_1q_1 + p_2q_2)\sqrt{2H_1}\\
\\
&+&(3\kappa^2 + 4\kappa q_1 + q_1^2 + q_2^2 + r_1^2)p_1^2 - 2\kappa p_1p_2q_2 + (\kappa^2 - 2\kappa q_1 + q_1^2 + q_2^2 + r_1^2)p_2^2\,,\\
\\
\Psi_+^{(1)}&=&-\kappa\alpha(\kappa + 2q_1)-\frac{ \kappa(\kappa r_2 + 2q_1r_1)\beta}{r_2}- 2\kappa(\kappa p_1q_1 + p_1q_1^2 + p_1r_1^2 + p_2q_1q_2)\sqrt{2H_1}\\
\\
&-& 2\kappa r_1(\kappa p_1^2 + 2p_1^2q_1 + p_1p_2q_2 + p_2^2q_1)\,,\\
\\
\Psi_+^{(0)}&=&\frac{\kappa^2(4q_1^2 + r_1^2 - r_2^2)\alpha}{4r_1}+\frac{\kappa^2(4q_1^2 - r_1^2 + r_2^2)\beta}{4r_2}
 + 2\kappa^2p_1q_1r_1\sqrt{2H_1} \\
 \\
 &-&\kappa^2p_1(\kappa^2p_1 - p_1q_1^2 + p_1q_2^2 - p_1r_1^2 - 2p_2q_1q_2)\,.
\end{array}
\]
Transformation $\beta\to -\beta$ in the $\Psi_+^{(k)}$ yields coefficients $\Psi_-^{(k)}$ of the Abel's polynomial $\Psi_-(\chi)$ on $E_-$.

The corresponding Lax matrices $L_{\pm}(\chi)$ (\ref{lax-3}) have the standard form \cite{be90,in07,ts04}:
\bq\label{lax-33}
L_{\pm}(\chi)=\left(
        \begin{array}{cc}
         V_R(\chi) & U_{\pm}(\chi) \\ \\
         \dfrac{ (\chi^2-\kappa^2)\bigl(2H_1\chi^2-(\alpha\pm\beta)\chi-H_2\bigr)-V_R^2(\chi)}{U_\pm(\chi)}\quad & -V_R(\chi) \\
        \end{array}
       \right)\,,
\eq
where
\[
V_R(\chi)= V(\chi)+R_\pm(\chi) U_{\pm}(\chi) \,,
\]
and
\[\begin{array}{rcl}
U_+(\chi)&=&\mbox{MakeMonic}\,\Psi_+=(\chi-\chi_1)(\chi-\chi_3)(\chi-\chi_5)\,,\qquad \chi_1=u_1\,,\\
\\
U_-(\chi)&=&\mbox{MakeMonic}\,\Psi_-=(\chi-\chi_2)(\chi-\chi_4)(\chi-\chi_6)\,,\qquad \chi_2=u_2\,.
\end{array}
\]
Here elliptic coordinates $\chi_{1,2}=u_{1,2}$ and functions $\chi_3,\ldots \chi_6$ on the phase space
are abscissas of the six intersection points $P_1,\ldots,P_6$ of the curves $E_\pm$ (\ref{epm-4}) with parabola $\Upsilon$ (\ref{par}).

Second Lax matrices $M_\pm$ have the form (\ref{mlax-3})
\[
M_\pm(\chi)=\left(
    \begin{array}{cc}
     0 & \dfrac{1}{2V_R}\,\dfrac{d}{dt} U_\pm \\ \\
     \dfrac{1}{U_\pm} \,\dfrac{d}{dt}V_R + \dfrac{W_R}{2V_RU_\pm}\,\dfrac{d}{dt}U_\pm & 0 \\
    \end{array}
   \right)(\chi)\,,
\]
where time $t$ is defined by equation (\ref{eq-ab-t}).

Semi-reduced divisors $\widetilde{D}_+=P_3+P_5$ and $\widetilde{D}_-=P_4+P_6$ of degree two have the following Mumford's coordinates
\[V(\chi)
=\sqrt{2H_1}\chi^2-\Bigl(p_1(\kappa + q_1)+ p_2q_2 +\sqrt{2H_1} r_1\Bigr)\chi+\kappa\left(q_1-\sqrt{2H_1} + p_1r_1\right)
\]
and
\bq\label{utx-car}
\widetilde{U}_\pm(\chi)=\chi^2+\left(\frac{r_1+r_2}{2} + \frac{\Psi_\pm^{(2)}}{\Psi_{\pm}^{(3)}}\right) \chi+
\left(\frac{(r_1+r_2)^2}{4} + \frac{(r_1+r_2)}{2}\,\frac{\Psi_\pm^{(2)}}{\Psi_\pm^{(3)}}+ \frac{\Psi_\pm^{(1)}}{\Psi_\pm^{(3)}}\right)\,.
\eq
It allows us to construct Lax matrices
\[
\widetilde{L}_{\pm}(\chi)=\left(
        \begin{array}{cc}
         V+R_\pm\widetilde{U}_{\pm} & \widetilde{U}_{\pm} \\ \\
         \dfrac{ (\chi^2-\kappa^2)\bigl(2H_1\chi^2-(\alpha\pm\beta)\chi-H_2\bigr)-\Bigl(V+R_\pm\widetilde{U}_{\pm}\Bigr)^2}{ \widetilde{U}_\pm}\quad & -V-R_\pm\widetilde{U}_{\pm} \\
        \end{array}
       \right)
\]
with the same spectral curves $E_{\pm}$ and the corresponding second matrices $\widetilde{M}_\pm$ of the form (\ref{mlax-3}).
The corresponding Baker-Akhieser functions $\psi_\pm$ and $\widetilde{\psi}_\pm$ have the tree and two poles on the common spectral curves $E_{\pm}$, respectively.

\section{Conclusion}
In classical mechanics, the Hamiltonian  equations on $2n$-dimensional phase space can be written as a Lax equation
\[
\frac{d}{dt} L(x)=[ L(x),M(x)]\,,
\]
 with the Lax matrix $ L(x)$ having a spectral curve of genus $g$ up to trivial coverings. The Lax matrices with $n>g$ are known for some St\"{a}ckel systems \cite{ee94}, Kowalevski top \cite{brs99}, Clebsch system \cite{p81}, Euler tops on $so(n)$ \cite{hei84}, classical magnets \cite{skl86}, etc.

 When the number of degrees of freedom $n$ is more than genus $g$ of the spectral curve we can suppose that:
 \begin{itemize}
  \item if $n>g$, we have superintegrable Hamiltonian system with $n-g$ additional independent integrals of motion, according to the Riemann-Roch theorem \cite{ts20r,ts20};
  \item if $n>g$, we have Hamiltonian system integrable by Abel's quadratures on the $n$-dimensional reducible Abelian varieties
\[A\cong A_1\times\cdots\times A_k,\]
and spectral curve of the corresponding Lax matrix $L(x)$ is one of the factors $A_i$ so that \[\mbox{dim}A=n>\mbox{dim}A_i=g.\]
 \end{itemize}
\par\noindent
In this note, we present Lax matrices for Euler's two centers problem at $n=2>g=1$.

Let us now discuss properties of the corresponding Baker-Akhiezer functions $\psi_{\pm}$. Abscissas of poles $P_i=(x_i,y_i)$, $i=1,\ldots,6$ are roots of the cubic Jacobi's polynomials
(\ref{ux}) or (\ref{ux-car})
\[
U_+=(x-x_1)(x-x_3)(x-x_5)\,,\quad\mbox{or}\quad U_-=(x-x_2)(x-x_4)(x-x_6)
\]
which are the first Mumford's coordinates of divisors $D_\pm$ on the elliptic curves $E_\pm$.

For Euler's system poles of the Baker-Akhiezer functions or roots of polynomials $U_\pm(x)$ have the different properties:
\begin{itemize}
\item abscissas $x_1$ and $x_2$ are simple functions (\ref{x12}) on elliptic coordinates (\ref{ell-coord}), that allows as to express original variables $p_{1,2}$ and $q_{1,2}$ in term of the Weierstrass $\wp$-function \cite{bis16};
\item other two roots $x_3,x_5$ or $x_4,x_6$ of polynomial (\ref{ux-p},\ref{utx-car}) are more complicated algebraic functions on elliptic coordinates (\ref{ell-coord}) and momenta (\ref{ell-mom}) and, therefore, expressions for original variables via these roots do not allow us to say anything definite about the properties of these coordinates us functions on time.
\end{itemize}
Thus, if we want to get explicit expressions for original variables we have to develop an algorithm for a search of the Baker-Akhiezer function poles which have the simplest relations with the original variables on the phase space.

\end{document}